\documentclass[aps,prl,twocolumn,superscriptaddress]{revtex4}
\usepackage{graphics}
\usepackage{graphicx}
\usepackage{amssymb}
\usepackage{hyperref}
\usepackage{color}

\newcommand{\fesige}{FeSi$_{1-x}$Ge$_x$ }

\begin{document}

\title{Metallic State in Cubic FeGe beyond its Quantum Phase Transition}
\author{P. Pedrazzini}
\affiliation{DPMC, University of Geneva, 24 Quai Ernest-Ansermet,
1211 Gen\`eve 4, Switzerland}

\author{H. Wilhelm}\thanks{Present address: Diamond Light Source Limited,
Chilton, Didcot, Oxfordshire, OX11 0DE, United Kingdom}

\affiliation{Max Planck Institute for Chemical Physics of Solids,
01187 Dresden, Germany}

\author{D. Jaccard}
\affiliation{DPMC, University of Geneva, 24 Quai Ernest-Ansermet, 1211 Gen\`eve 4, Switzerland}

\author{T. Jarlborg}
\affiliation{DPMC, University of Geneva, 24 Quai Ernest-Ansermet, 1211 Gen\`eve 4, Switzerland}

\author{M. Schmidt}
\affiliation{Max Planck Institute for Chemical Physics of Solids,
01187 Dresden, Germany}

\author{M. Hanfland}
\affiliation{European Synchrotron Radiation Facility, 38043
Grenoble, Cedex, France}

\author{L. Akselrud}
\affiliation{Lviv State University, Lviv, Ukraine}

\author{H. Q. Yuan}\thanks{Present address: Department of Physics, University of
Illinois, Urbana, IL 61801, USA} \affiliation{Max Planck Institute
for Chemical Physics of Solids, 01187 Dresden, Germany}

\author{U. Schwarz}
\affiliation{Max Planck Institute for Chemical Physics of Solids,
01187 Dresden, Germany}

\author{Yu. Grin}
\affiliation{Max Planck Institute for Chemical Physics of Solids,
01187 Dresden, Germany}

\author{F. Steglich}
\affiliation{Max Planck Institute for Chemical Physics of Solids,
01187 Dresden, Germany}

\begin{abstract}
We report on results of electrical resistivity and structural
investigations on the cubic modification of FeGe under high
pressure. The long-wavelength helical order ($T_C=280$~K) is
suppressed at a critical pressure $p_c\approx 19$~GPa. An anomaly in
the resistivity data at $T_X(p)$ and strong deviations from a
Fermi-liquid behavior in a wide pressure range above $p_c$ suggest
that the suppression of $T_C$ disagrees with the standard notion of
a quantum critical phase transition. The metallic ground state
persisting at high pressure can be described by band-structure
calculations if structural disorder due to zero-point motion is
included. Discontinuous changes in the pressure dependence of the
shortest Fe-Ge interatomic distance occurring close to the $T_C(p)$
phase line could be interpreted as a symmetry-conserving transition
of first order.
\end{abstract}



\maketitle

The electronic and magnetic properties of binary compounds
crystallizing in the $B20$ structure, such as the monosilicides of
Mn, Cr, Fe, and Co are an active topic of research in condensed
matter physics. Among them, FeSi and MnSi have attracted renewed
interest. FeSi is a semiconductor with a narrow gap $E_g\approx
80\,$meV and due to its peculiar magnetic and optical properties at
low temperature, $T$, it is sometimes referred to as Kondo insulator
or correlated insulator \cite{Wertheim65,Schlesinger93}. Band
structure calculations reproduce the gap as well as the recently
observed transition to a metallic phase in \fesige at a critical
concentration $x_{\rm c}\approx 0.25$
\cite{Yeo03,Anisimov02,Jarlborg04}. Accordingly, it was evident that
FeGe is a rare case where external pressure, $p$, might induce a
symmetry-retaining transition from a metallic to an insulating
state.

MnSi and FeGe are prominent examples where the Dzyaloshinskii-Moriya
interaction causes a modulation of the ferromagnetic structure as a
consequence of the lack of inversion symmetry in the $B20$ structure
(space group $P2_13$) \cite{Bak80}. In MnSi, the helical order
occurs below $T_{\rm C}=29\,$K. The modulation has a wavelength of
$175\,$\AA\ and the ordered moments of about $m=0.4\,\mu_B$ (with
$\mu_B$ the Bohr magneton) per Mn atom are perpendicular to the
spiral propagation vector ${\bf k}\parallel [1 1 1]$. It is well
established that the second order phase transition is driven first
order for a sufficiently weak magnetic interaction close to the
critical pressure, $p_{\rm c}=1.46\,$GPa
\cite{Vojta01,Fak05,Pfleiderer04,Thessieu95}. In a wide $p$-range
above $p_{\rm c}$, MnSi presents unusual physical properties, such
as non-Fermi liquid (NFL) behavior in the electrical resistivity,
$\rho(T)\propto T^{3/2}$
\cite{Thessieu95,Pfleiderer97,Pedrazzini05}, or partial ordering
suggesting a magnetic state at high $p$ \cite{Pfleiderer04}. In
FeGe, on the other hand, helimagnetism sets in through a first order
phase transition at $T_{\rm C}=280\,$K with a saturated moment of
$m=1\mu_B$ per Fe atom \cite{Waeppling68}. The helical modulation
has a period of about $700\,$\AA\ and propagates along $[1 0 0]$. It
alters its direction to ${\bf k}\parallel [1 1 1]$ at $T_2\approx
211-245\,$K without a change in the period \cite{Lebech89}. Given
the structural and magnetic similarities between MnSi and FeGe, it
seems very likely that a volume compression in FeGe could eventually
suppress the long-range magnetic order and reveal strong deviations
from the standard notion of a Landau-Fermi liquid (LFL) or even,
recalling the metal-to-insulator transition (MIT) in
FeSi$_{1-x}$Ge$_x$, induce an insulating state.

In this Letter, we explore the $(T,p)$ phase diagram of cubic FeGe
by means of electrical resistivity, $\rho(T)$, and angle-dispersive
X-ray diffraction experiments. The $\rho(T)$ results point to a
suppression of the helical order at $p\approx 19\,$GPa while a
metallic ground state with unusual low-$T$ transport properties
persists up to 23~GPa, the highest $p$ achieved in this
investigation. Our band-structure calculations suggest that disorder
due to zero-point motion (ZPM) is strong enough to close the narrow
gap expected for compressed FeGe, stabilizing a new magnetic ground
state. An anomaly at a temperature $T_X$ observed above $p_c$ might
be related to this magnetic phase. The isothermal structural data at
low $T$ provide some indications of a symmetry-conversing
first-order phase transition close to the $T_C(p)$ phase boundary.

Single crystals of cubic FeGe were grown by vapor transport in a
two-zone furnace, using iodine as chemical agent
\cite{Richardson67}. FeGe crystallized very slowly by an endothermal
transport reaction from 850~K to 810~K. The largest pieces were
examined thoroughly by various X-ray techniques and electron-beam
microanalysis. Four-probe $\rho(T)$ measurements were carried out on
a $690\times 105\times 23\,\mu{\rm m}^3$ parallelepiped with the
current ($4\le j\le 80\,{\rm A\,cm}^{-2}$) applied perpendicular to
a $[111]$ direction. The sample was mounted together with a strip of
Pb, which served as $p$-gauge, in a Bridgman-type high-pressure cell
using steatite as $p$-transmitting medium \cite{Jaccard98}.
Additional low-$p$ data were obtained using a second pressure cell.
The X-ray diffraction experiments were performed on well ground and
annealed powder (at 670 K for two days) using a diamond-anvil cell
with helium as $p$-medium and the fluorescence peaks of
SrB$_4$O$_7$:Sm$^{2+}$ as $p$-gauge \cite{Datchi97}.

\begin{figure}
\begin{center}
\includegraphics[width=0.45\textwidth]{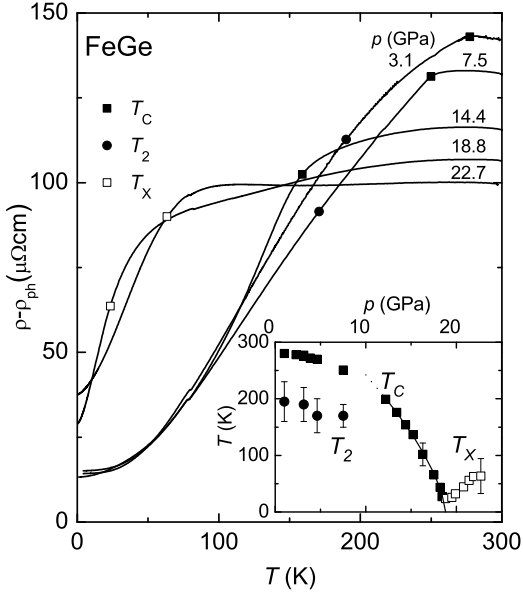}
\caption{Non-phononic contribution to the electrical resistivity in
cubic FeGe, $\rho-\rho_{\rm ph}$. The onset of different spiral
magnetic ordering at $T_{\rm C}$ and $T_2$ leads to anomalies in
$\rho(T)$, indicated by the symbols. The nature of the anomaly at
$T_X$ is not known. Inset: $(T,p)$ phase diagram of cubic FeGe. The
full line is a fit to $T_{\rm C}(p)$ which yields $p_{\rm
c}=18.8\,$GPa (see text).\label{fig:rhovst}}
\end{center}
\end{figure}

Figure~\ref{fig:rhovst} shows the $T$-dependence of the non-phononic
contribution, $\rho(T)-\rho_{\rm ph}(T)$, of cubic FeGe. The phonon
contribution, $\rho_{\rm ph}(T)$, is described with the
Bloch-Gr\"uneisen formula using both $p$-independent Debye
temperature, $\Theta_{\rm D}=250\,$K, and amplitude, $\rho_{\rm
ph}(300\,{\rm K})=87\,\mu\Omega\,{\rm cm}$. For $p<12\,$GPa the
sharp kink is attributed to $T_C$. An additional broad anomaly is
resolved at $T_2$ up to $8\,$GPa. It seems to be associated with the
sluggish change in the modulation vector observed at $p=0$
\cite{Lebech89}. Above $p\approx 12\,$GPa, only $T_C$ is detected
and it decreases by an order of magnitude upon approaching
$18.4\,$GPa (inset to Fig.~\ref{fig:rhovst}). Slightly above this
$p$, however, the slope change in $\rho(T)$ occurs at higher $T$ and
therefore we label it $T_X$. Surprisingly, $T_X$ increases steadily
with $p$. We want to stress that at $p$ as high as $19\,$GPa,
unavoidable strain or $p$-gradients could obscure two adjacent
anomalies in $\rho(T)$ and impede the precise detection of $T_C(p)$
below $25\,$K \cite{LeadWidth}. Although the residual resistivity
has doubled upon reaching 23~GPa, FeGe remains metallic. For
$p>12\,$GPa, $\rho(T)$ was measured down to $50\,$mK with low
current densities and no hints of superconductivity were found. The
$(T,p)$ phase diagram is depicted in the inset to
Fig.~\ref{fig:rhovst}. Fitting $T_C(p)=T_C(0)(1-p/p_c)^{\nu}$ to the
data for 12.2~GPa$\,\leq p \leq 18.4\,$GPa yields
$T_C(0)=400(20)\,$K, $p_c=18.8(1)\,$GPa, and the exponent
$\nu=0.66(4)$. The latter is almost identical to the exponent found
in MnSi \cite{Thessieu95}. Thus, by analogy, long-range
helimagnetism in FeGe seems to be suppressed at $p_c=18.8$~GPa. The
origin of $T_X$ cannot be inferred from $\rho(T)$ experiments,
though a magnetic nature seems very likely. This notion is
encouraged by the observation, that for $T>T_X$, as well as above
$T_C$, the non-phononic electrical resistivity is almost
$T$-independent, as expected for spin-disorder resistivity.

\begin{figure}
\begin{center}
\includegraphics[width=0.45\textwidth]{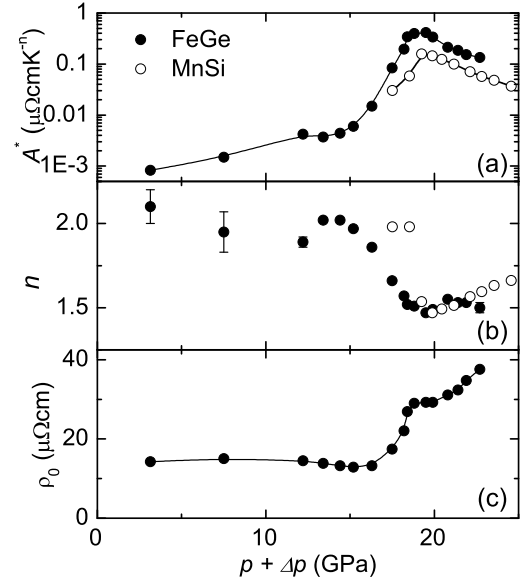}
\caption{Pressure-dependence of the resistivity parameters $A^*$,
$n$, and $\rho_0$ in cubic FeGe. For comparison, MnSi data for $A^*$
and $n$ taken from Ref.~\cite{Pedrazzini05} are included but shifted
by $\Delta p=17.3\,$GPa to match $p_c$ of FeGe.
\label{fig:rhoparameter}}
\end{center}
\end{figure}

The $T$-variation of $\rho(T)$ for $T<10\,$K is described by
$\rho(T)=\rho_0+A^* T^n$, with $\rho_0$ the residual resistivity and
$A^* T^n$ accounting for electron-magnon and/or electron-electron
scattering. Figure \ref{fig:rhoparameter} shows that $\rho_0\approx
14\,\mu\Omega\,{\rm cm}$ and the exponent $n\approx 2$ remain almost
unchanged below $p\approx 15\,$GPa, while the slow increase of
$A^*(p)$ could reflect the enhanced electron-magnon scattering. For
$p>15\,$GPa, however, all parameters change drastically, and
$A^*(p)$ as well as $n(p)$ show striking similarities to MnSi close
to its zero-$T$ phase transition (open symbols in
Fig.~\ref{fig:rhoparameter}). Given the giant $A^*(p)$ maximum at
$p_c$ and the particular values for the exponents $\nu=0.66$ and
$n\approx 3/2$, it is tempting to associate $p_c$ with a quantum
critical point (QCP). The value $A^*(p_c)\approx 0.4\,\mu\Omega{\rm
cm\,K}^{-1.5}$ is much larger than those measured in other 3d/4d
itinerant ferromagnets close to a QCP, like Ni$_3$Al
\cite{Niklowitz05} or Ni$_x$Pd$_{1-x}$ \cite{Nicklas99}. It is even
bigger than $A^*(p_c)$ in MnSi despite the larger $p$-gradient
expected at higher $p$. In a QCP scenario one would expect a narrow
crossover regime where the LFL description breaks down. In FeGe,
however, the exponent $n\approx 3/2$ remains $p$-independent up to
the highest $p$ while $A^*(p)$ decreases rapidly. The strong
increase of $\rho_0$ upon approaching $p_c$
(Fig.~\ref{fig:rhoparameter}(c)) can be interpreted as an enhanced
impurity scattering due to ferromagnetic quantum fluctuations
\cite{Miyake02} whereas strong magnetic disorder or a near-lying
insulating phase could be responsible for the steady increase of
$\rho_0(p)$ above $p_c$. The similarity of our observations with
those in MnSi \cite{Pfleiderer97,Thessieu95,Doiron03} is striking
and FeGe seems to be another candidate for the breakdown of the LFL
model with a puzzling quantum phase of matter above $p_c$
\cite{Doiron03}.

\begin{figure}
\begin{center}
\includegraphics[width=0.5\textwidth]{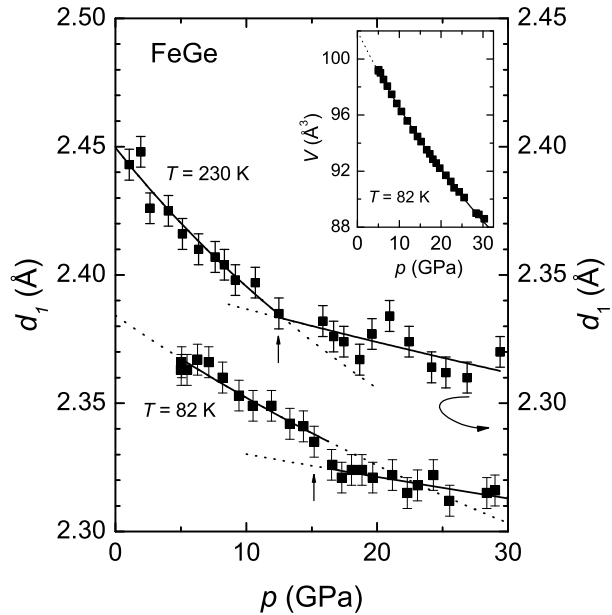}
\caption{Pressure dependence of the Fe-Ge distance $d_1$ (pointing
along [111]) at 82~K (left scale) and 230~K (right scale) in cubic
FeGe. Lines are fits to the data and error bars represent three
times the standard deviation. The vertical arrows indicate the onset
of the anomaly in $d_1(p)$. Inset: $V(p)$ data and fit for
$T=82\,$K. \label{fig:distances82k}}
\end{center}
\end{figure}

\begin{figure}
\begin{center}
\includegraphics[width=0.45\textwidth]{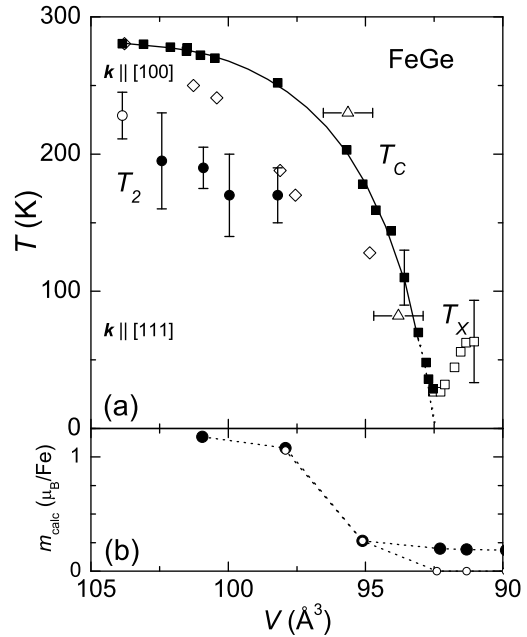}
\caption{(a) $(T,V)$ phase diagram of cubic FeGe obtained from
$\rho(T)$ and structural data (triangles). The $T_2$ data point
(circle) at zero compression is taken from Ref.~\cite{Lebech89}.
$V(x)$ data on FeSi$_{1-x}$Ge$_x$ (diamonds) are taken from
Ref.~\cite{Yeo03} and are corrected for thermal contraction. (b)
Volume dependence of the calculated magnetization $m_{\rm calc}$
with and without zero-point motion (bold and open symbols,
respectively).\label{fig:phasediagram}}
\end{center}
\end{figure}

The analysis of the low-$T$ powder diffraction data provide some
indications of a symmetry-conserving first-order transition. Figure
\ref{fig:distances82k} shows the shortest interatomic Fe-Ge distance
along the $[111]$ direction, $d_1$, as obtained by a full profile
refinements of the diffraction pattern. The low-$p$ $d_1(p)$
behavior for $82\,$K and $230\,$K described by an appropriate
equation-of-state (EOS) clearly fails to account for the
$p$-dependence above $15\,$GPa and $12\,$GPa, respectively. The two
remaining Fe-Ge distances as well as all distances at room
temperature decrease smoothly with $p$. The anomalies in $d_1(p)$
agree remarkably well with the $T_C(p)$ phase boundary deduced from
the $\rho(T)$ data (inset to Fig.~\ref{fig:rhovst}) and pose the
question about the nature of the phase transition observed in
resistivity. From a structural point of view they may mark the start
of a lattice instability such as an incipient martensitic
transition. Further investigations are necessary to clarify this
point. The unit-cell volume, $V$, decreases continuously for all
investigated $T$. As an example, data for $T=82$~K are shown in the
inset to Fig.~\ref{fig:distances82k}. $V(p)$ can be described by a
Murnaghan EOS \cite{Murnaghan44}, using a bulk modulus
$B_0=147(3)\,$GPa, its $p$-derivative $B'=4.4(2)$ and
$V(0)=102.43(7)\,{\rm \AA}^3$ \cite{vol_vs_p}.

The structural information obtained at different $T$ allows the
$(T,p)$ data (inset to Fig.~\ref{fig:rhovst}) to be converted into
the $(T,V)$ phase diagram presented in
Fig.~\ref{fig:phasediagram}(a). The change of compressibility with
$T$ was considered \cite{vol_vs_p}. The anomalies in the Fe-Ge
distances are also added. They agree fairly well with the $T_C(V)$
phase boundary deduced from the $\rho(T)$ data. This accordance is
remarkable if the quite different $p$ environments are recalled
(steatite vs. helium). The smooth evolution of $T_{\rm C}(V(p))$ in
FeGe resembles the one observed in MnSi. In both compounds $T_{\rm
C}(V)\rightarrow 0$ at a similar volume $V_{\rm c}\approx 93\,{\rm
\AA}^3$, despite the subtle difference in the magnetic modulation,
and NFL behavior is observed over a wide range of lattice
compression. To our knowledge only one other metallic lattice, pure
Fe, shows a similar behavior. In this case, a connection between
correlated electron phenomena, including superconductivity, and the
martensitic transition driven by magnetism has been considered
\cite{Holmes04}.

The $T_C(V(x))$ data of \fesige \cite{Fisk} are also included in
Fig.~\ref{fig:phasediagram}. Although the initial suppression of
$T_C$ is stronger in \fesige than in FeGe, both systems have a
similar $T_C$ at $V\approx 95\,{\rm \AA}^3$. At larger compression,
however, a striking difference occurs: In \fesige long-range
magnetic order disappears at a first order MIT (at $V(x_c)\approx
95$\,\AA$^3$ \cite{Fisk}) whereas $T_C(V)$ in FeGe still continuous
to decrease. Furthermore, in FeGe, a metallic ground-state is
observed down to a compression comparable to the unit-cell volume of
semiconducting FeSi at $p=0$ ($V\approx 90\,{\rm \AA}^3$).

We have addressed this unexpected result by performing
self-consistent spin-polarized linear muffin-tin orbital (LMTO)
calculations for lattice constants in the range $4.65\,{\rm
\AA}<a<4.08\,$\AA, corresponding to a calculated $p$-range $5\,{\rm
GPa}<p<180\,$GPa. The first set of results in Ref.~\cite{Jarlborg04}
were done without considering the effect of ZPM. In the paramagnetic
calculations the gap was found to increase steadily from
$E_g\approx15\,$meV at the largest $V$ to $E_g\approx 35\,$meV at
the smallest one, compared to $E_g\approx80\,$meV for FeSi
\cite{Jarlborg99}. The resulting magnetic moment per Fe atom,
$m_{\rm calc}$, follows rather closely the reduction of the measured
$T_C(V)$ of FeGe (open symbols in Fig.~\ref{fig:phasediagram}(b))
and suggests a loss of magnetism around $92\,{\rm \AA}^3$, in
contrast to our observations.

However, if structural disorder due to ZPM is considered, $m_{\rm
calc}$ remains finite in the $p$-range probed by our experiment. The
calculations were made for a 64-atom cell in which the atomic
positions deviate randomly from their ideal ones, as described for
thermal disorder in FeSi \cite{Jarlborg99}. The mean deviation
$u(T)=0.03$\,\AA, for $T \to 0$, was used for Fe and Ge, which is
about 40\% of the value used for FeSi \cite{Jarlborg99}. In the case
of FeGe at $a=4.61\,$\AA\ (i.e.~$V\approx 98$\,\AA$^3$), when the
system is metallic and strongly magnetic, there are only minor
changes caused by the small structural disorder caused by the ZPM.
The density of states (DOS) is large at the Fermi energy and ZPM has
very little effect on $m_{\rm calc}$ (see
Fig.~\ref{fig:phasediagram}(b)). But for $V\lesssim 92.5\,{\rm
\AA}^3$, when $E_g\approx 27$~meV in the perfect structure, the ZPM
is able to close the gap and cause magnetism. The small
averaged-moments, $m_{\rm calc}\approx 0.18-0.08\,\mu_{\rm B}$,
persist until a compression to $V\approx 83\,{\rm \AA}^3$ which
corresponds to $p\approx 40-50$~GPa. The moments are not uniformly
distributed on the Fe atoms but can vary by a factor of four among
the different Fe-sites.

The main difference between band-structure calculations for FeGe
with and without ZPM is that the gap has disappeared in the former
case. Therefore, the persistence of magnetism at large compression
can be the result of ZPM, which tends to push the MIT to higher $p$.
Nevertheless, the nature of such magnetism should be different from
that in clearly metallic high-DOS systems: Since the structure is
continuously changing on a short time scale, it is expected that
also the local moment on each Fe will change in time. For even
larger lattice densities, $E_g$ will widen further and approach the
situation in FeSi, where ZPM has no effect. These results provide no
immediate explanation to why $T_X$ could increase with $p$. On the
other hand, a QCP, which separates paramagnetism and Stoner
magnetism, can be very sharp in calculations for perfect lattice
with diverging exchange enhancement. But disorder due to ZPM will
smear the transition and $p_c$ marks the changes from static to
dynamically changing magnetic moments.

In conclusion, the electrical resistivity and structural data
provide evidence for the suppression of the long-range magnetic
order in cubic FeGe at a critical pressure $p_c\approx 19$~GPa
though at higher $p$ an anomaly at $T_X$ might indicate some kind of
magnetic correlations. The abrupt changes in the $p$-dependence of
the shortest Fe-Ge distance could be the trace of a
symmetry-retaining first-order phase transition which is in good
agreement with the $T_C(p)$ phase boundary. The metallic state,
although with unusual properties, persists up to 23~GPa, the maximum
$p$ achieved. The wide $p$-range of the non-Fermi liquid behavior in
connection with the exponent $n\approx 3/2$ and the giant value of
the temperature coefficient, $A^*$, is interpreted as a breakdown of
the standard scenario of a quantum critical phase transition. The
band-structure calculations show that zero-point motion can overcome
the narrow gap expected in FeGe at moderate compression. A
semiconducting ground state, however, is predicted at much larger
lattice densities.

\begin{acknowledgments}
We acknowledge U. Burkhardt, R. Cardoso, Yu. Protz, R. Ramlau, W.
Schnelle, and H. Zhang for cutting and characterizing the crystals.
We are grateful to R. Demchneya and A. Wosylus for their commitment
during the X-ray experiments and to Z. Fisk for enlightening
discussions.
\end{acknowledgments}


\begin{thebibliography}{99}

\bibitem{Wertheim65} G. K. Wertheim \emph{et al.}, Phys. Lett. {\bf 18}, 89 (1965).

\bibitem{Schlesinger93} Z. Schlesinger \emph{et al.}, Phys. Rev. Lett. \textbf{71}, 1748
(1993).

\bibitem{Yeo03} S. Yeo {\ et al.}, Phys. Rev. Lett. {\bf 91}, 046401 (2003).

\bibitem{Anisimov02} V. I. Anisimov {\em et al.}, Phys. Rev. Lett. {\bf 89}, 257203 (2002).

\bibitem{Jarlborg04} T. Jarlborg, J. Mag. Mag. Mater. {\bf 283}, 238 (2004).

\bibitem{Bak80} P. Bak and M. H. Jensen, J. Phys. C: Solid State Phys. {\bf 13}, L881 (1980).

\bibitem{Vojta01} T. Vojta and R. Sknepnek, Phys. Rev. B
\textbf{64}, 052404 (2001).

\bibitem{Fak05} B. F{\aa}k \emph{et al.}, J. Phys.: Condens. Matter \textbf{17}, 1635 (2005).

\bibitem{Pfleiderer04} C. Pfleiderer \emph{et al.}, Nature \textbf{427}, 227 (2004).

\bibitem{Thessieu95} C. Thessieu {\em et al.}, Solid State Commun.
\textbf{95}, 707 (1995); C. Thessieu, Ph.D. thesis, Universit\'e
Paris VII (1995).

\bibitem{Pfleiderer97} C. Pfleiderer \emph{et al.}, Phys. Rev. B \textbf{55}, 8330 (1997).

\bibitem{Pedrazzini05} P. Pedrazzini \emph{et al.}, Physica B
\textbf{378-380}, 165 (2006); cond-mat/0509772.

\bibitem{Waeppling68} R. W\"appling and L. H\"aggstr\"om, Phys.
Lett. A, \textbf{28}, 173 (1968).

\bibitem{Lebech89} B. Lebech, J. Bernhardt, and T. Freltoft, J. Phys.: Condens. Matter {\bf 1}, 6105 (1989).

\bibitem{Richardson67} M. Richardson, Acta Chem. Scand. {\bf 21}, 2305
  (1967).

\bibitem{Jaccard98} D. Jaccard \emph{et al.}, Rev. High Pressure Sci. Technol. {\bf 7}, 412 (1998).

\bibitem{Datchi97} F. Datchi, R. LeToullec, and P. Loubeyre, J.
Appl. Phys. \textbf{81}, 3333 (1997).

\bibitem{LeadWidth} We estimate a $p$-gradient $\Delta p\sim
0.5\,$GPa from the $10\%-90\%$ criterium for the superconducting
transition in Pb.

\bibitem{Niklowitz05} P. G. Niklowitz {\em et al.}, Phys. Rev. B
{\bf 72}, 024424 (2005).

\bibitem{Nicklas99} M. Nicklas {\em et al.}, Phys. Rev. Lett. {\bf 82}, 4268 (1999).

\bibitem{Miyake02} K. Miyake and O. Narikiyo, J. Phys. Soc. Jpn. {\bf 71}, 867 (2002).

\bibitem{Doiron03} N. Doiron-Leyraud {\em et al.}, Nature {\bf 425}, 595 (2003).

\bibitem{Murnaghan44} F. D. Murnaghan, Proc. Natl. Acad. Sci. USA {\bf 30}, 244 (1944).

\bibitem{vol_vs_p} The $V(p)$-dependence at $230\,$K [$290\,$K] is described  with
$B_0=135(1)\,$GPa [$130(1)\,$GPa], $B'=4.7(1)$, and
$V_0=103.17(3)\,{\rm \AA}^3$ [$103.86(2)\,{\rm \AA}^3$].

\bibitem{Holmes04} A. T. Holmes \emph{et. al}, J. Phys.: Condens.
Matter \textbf{16}, S1121 (2004).

\bibitem{Fisk} Z. Fisk, private communication.

\bibitem{Jarlborg99} T. Jarlborg, Phys. Rev. B {\bf 59}, 15002 (1999).

\end{thebibliography}
\end{document}